\documentclass[aip,jap,a4,reprint,twoside,floatfix,superscriptaddress]{revtex4-2}
                 
\usepackage{epsfig}                                                            
\usepackage{graphicx}
\usepackage{dcolumn}
\usepackage{color}
\usepackage{bm}
\usepackage{subfigure}
\usepackage{amsmath}

\newcommand{\be}{\begin{equation}} \newcommand{\ee}{\end{equation}}
\newcommand{\bea}{\begin{eqnarray}} \newcommand{\eea}{\end{eqnarray}}

\begin{document}

\title{\bf Enhanced negative capacitance in La-doped Pb(Zr$_{0.4}$Ti$_{0.6}$)O$_3$ ferroelectric capacitor from tuning of bias voltage pulse}

\author{Ganga S. Kumar} \affiliation{Multiscale Microstructure and Mechanics of Materials Division, CSIR-Central Glass and Ceramic Research Institute, Kolkata 700032 India} 
\author{Sudipta Goswami} \affiliation{School of Materials Science and Nanotechnology, Jadavpur University, Kolkata 700032, India}
\author{Subhashree Chatterjee}  \affiliation{School of Physical Sciences, Indian Association for the Cultivation of Science, Kolkata 700032, India} 
\author{Dilruba Hasina}  \affiliation{School of Physical Sciences, Indian Association for the Cultivation of Science, Kolkata 700032, India} 
\author{Miral Verma} \affiliation{School of Minerals, Metallurgical and Materials Engineering, Indian Institute of Technology, Bhubaneswar 752050, India} 
\author{Devajyoti Mukherjee} \affiliation{School of Physical Sciences, Indian Association for the Cultivation of Science, Kolkata 700032, India}
\author{Chandan Kumar Ghosh} \affiliation{School of Materials Science and Nanotechnology, Jadavpur University, Kolkata 700032, India}
\author{Dipten Bhattacharya} \email{dipten@cgcri.res.in} \affiliation{Multiscale Microstructure and Mechanics of Materials Division, CSIR-Central Glass and Ceramic Research Institute, Kolkata 700032 India} 

\date{\today}

\begin{abstract}
We report a remarkable bias voltage dependent specific negative capacitance in multidomain La-doped Pb(Zr$_{0.4}$Ti$_{0.6}$)O$_3$ (PLZT) ferroelectric capacitors. The specific negative capacitance maximizes at a specific bias voltage because of emergence of maximum domain-wall density during ``switching" of the domains. Domain configuration changes from such an ``optimum" state if higher or lower bias voltage is applied at a much faster or slower rate. Phase-field simulation using time-dependent Ginzburg-Landau equation corroborates the experimental results and shows dependence of the domain-wall length during switching on the bias voltage amplitude and its maximization at a specific bias voltage amplitude. Interestingly, the radius of curvature of the resulting polarization ($P$) versus voltage ($V$) hysteresis loop at the coercive voltage ($V_C$), as well, turns out to be depending on the bias voltage. All these results indicate a close correlation among the bias voltage pulse profile (amplitude and time scale), domain-wall length during switching, shape of the resulting ferroelectric hysteresis loop, and the transient negative capacitance. It may have important ramifications both in the context of physics behind negative capacitance in a multidomain ferroelectric capacitor and devices being developed by exploiting its advantages. 
\end{abstract}

\maketitle

\section{Introduction} 
A ferroelectric capacitor exhibits transient negative capacitance (TNC) during switching between two thermodynamically stable positive capacitive states of positive and negative saturation polarization ($\pm P_S$) \cite{Dutta,Iniguez}. For a monodomain system, the TNC results from mismatch between switching kinetics of bound ($\partial P/\partial t$) and free charge density ($\partial Q/\partial t$) at the ferroelectric layer and metal electrodes ($\partial Q/\partial t$$<$$\partial P/\partial t$), respectively \cite{Young}. In a multidomain system, on the other hand, it is governed by the difference in polarization and polarizability between the domain and domain wall \cite{Ramesh,Zubko}. This difference together with imperfect screening of charges at the domain wall results in a depolarizing field acting locally aganist the externally applied polarizing field. Charging of the multidomain capacitor is, therefore, associated with a voltage drop at a local scale which, in turn, yields negative capacitance. The domain-wall motion results in redistribution of this stray field at the interfaces of domain and domain wall which governs the magnitude of negative capacitance. Freezing of the negative capacitive state in such a multidomain system using a dielectric capacitor in series yields static negative capacitance (SNC). This, in turn, gives rise to substantial enhancement of the equivalent capacitance and hence opens vistas for many applications \cite{Alam}. Research in recent time has examined (albeit, primarily, theoretically) the influence of several factors on the TNC and SNC such as intrinsic viscosity of the ferroelectric, the domain switching kinetics (homogeneous or heterogeneous), the size effect (bulk versus nanoscale systems) etc. \cite{Hoffmann,Saha,Hwang,Bellaiche}. Interestingly, while TNC is examined in monodomain systems, SNC is considered to be relevant in a multidomain one \cite{Young}. However, whether TNC is relevant and how and why it is influenced by bias voltage pulse profile (amplitude and time scale) even in a multidomain system has not been investigated in detail. 

\begin{figure*}[ht!]
\centering
{\includegraphics[scale=0.40]{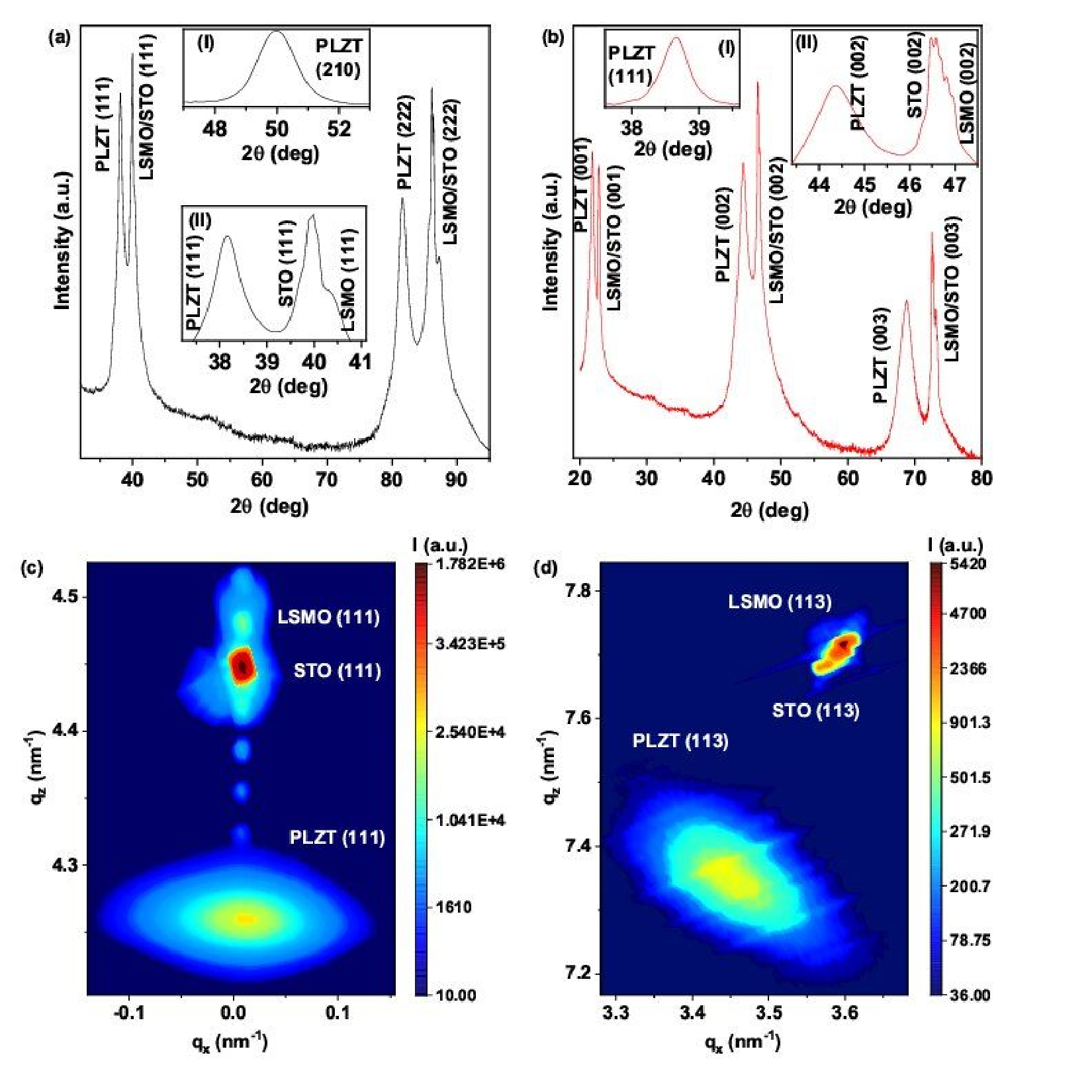}}
\caption{XRD $\theta-2\theta$ patterns of (a) PLZT/LSMO/STO (111) and (b) PLZT/LSMO/STO (001) thin film heterostructures. Insets show the detector scan of asymmetric and symmetric planes for the PLZT/LSMO/STO (111) and  PLZT/LSMO/STO (001) film. Reciprocal space mapping about (c) STO (111) symmetric plane of PLZT/LSMO/STO (111) and about (d) STO (113) asymmetric plane of PLZT/LSMO/STO (001) thin film heterostructures, respectively.} 
\end{figure*}
 
In this paper, we report significant dependence of specific transient negative capacitance on the amplitude and timescale of bias voltage pulse in a multidomain ferroelectric capacitor. The ferroelectric domain switching under a bias voltage pulse applied across a specific timescale (a bipolar pulse implemented within the Sawyer-Tower circuit) is governed by both nucleation and growth and free energy landscape of the ferroelectric system. While the free energy landscape is determined by the crystallographic structure of the domains as well as domain types (if different types of domains with 90$^o$ and 180$^o$ domain walls are present), defects in the system act as nucleation centers and influence the free energy landscape further. Recent work has shown \cite{Li} that indeed nucleation and growth together with free energy landscape influence the domain switching. The transient negative capacitance maximizes when the bias voltage pulse amplitude and timescale match the characteristic free energy landscape and thereby induce domain switching following the characteristic switching kinetics. It yields maximum domain wall density during switching - not achievable if higher or lower voltage is applied at a much faster or slower rate - because of completion of the nucleation and growth of reverse domains which, in turn, results in maximum transient negative capacitance. Using time-dependent Ginzburg-Landau equation, the domain switching kinetics has been simulated theoretically to examine how the domain-wall length evolves under different conditions - low and high bias voltage amplitude and low and high defect, i.e., nucleation center density. We found that indeed application of higher bias voltage results in faster switching yet smaller domain-wall area during switching because of prevalence of the unswitched states. System with higher defect density too yields similar results even though the timescale of switching turns out to be larger. These results indicate that higher defect density slows down the switching kinetics while for a system with a fixed defect density higher bias volatge results in incomplete nucleation and thus the switching states are dominated by the unswitched states. Interestingly, the radius of curvature of the resulting ferroelectric hysteresis loop too, turns out to be dependent on the bias voltage pulse amplitude and hence the domain switching pathway followed. Our work thus highlights a correlation among the bias voltage pulse amplitude and timescale, domain-wall length during switching, shape of the resulting ferroelectric hysteresis loop, and the transient negative capacitance. 

\begin{figure*}[ht!]

{\includegraphics[scale=0.65]{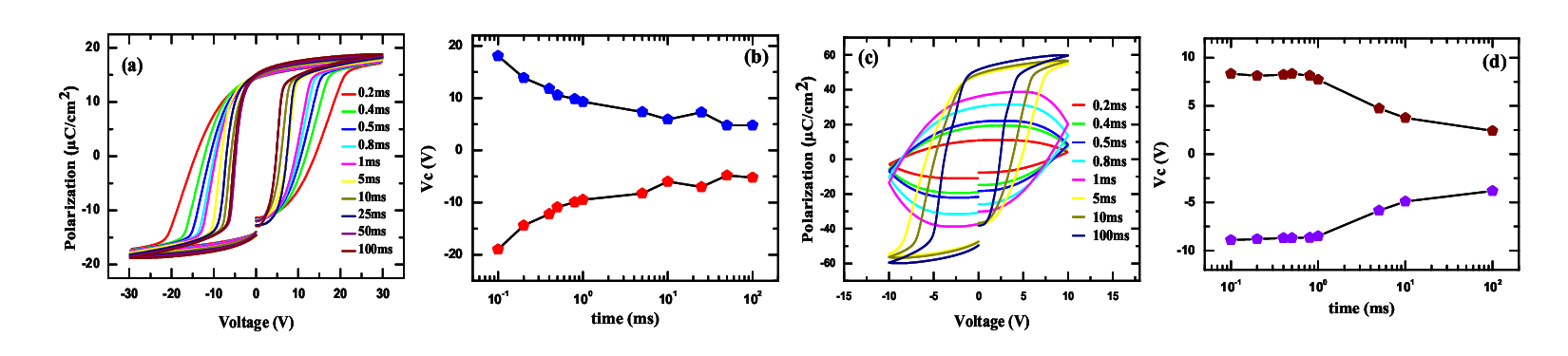}}
\caption{(a), (c) The polarization ($P$) versus bias voltage ($V$) ferroelectric hysteresis loops recorded by using standard triangular bipolar voltage pulse of different time scale for, respectively (111) and (001) films; (b), (d) corresponding variation of the $V_C$ with time scale of the bias voltage pulse. }
\end{figure*}     

\begin{figure}[ht!]

{\includegraphics[scale=0.30]{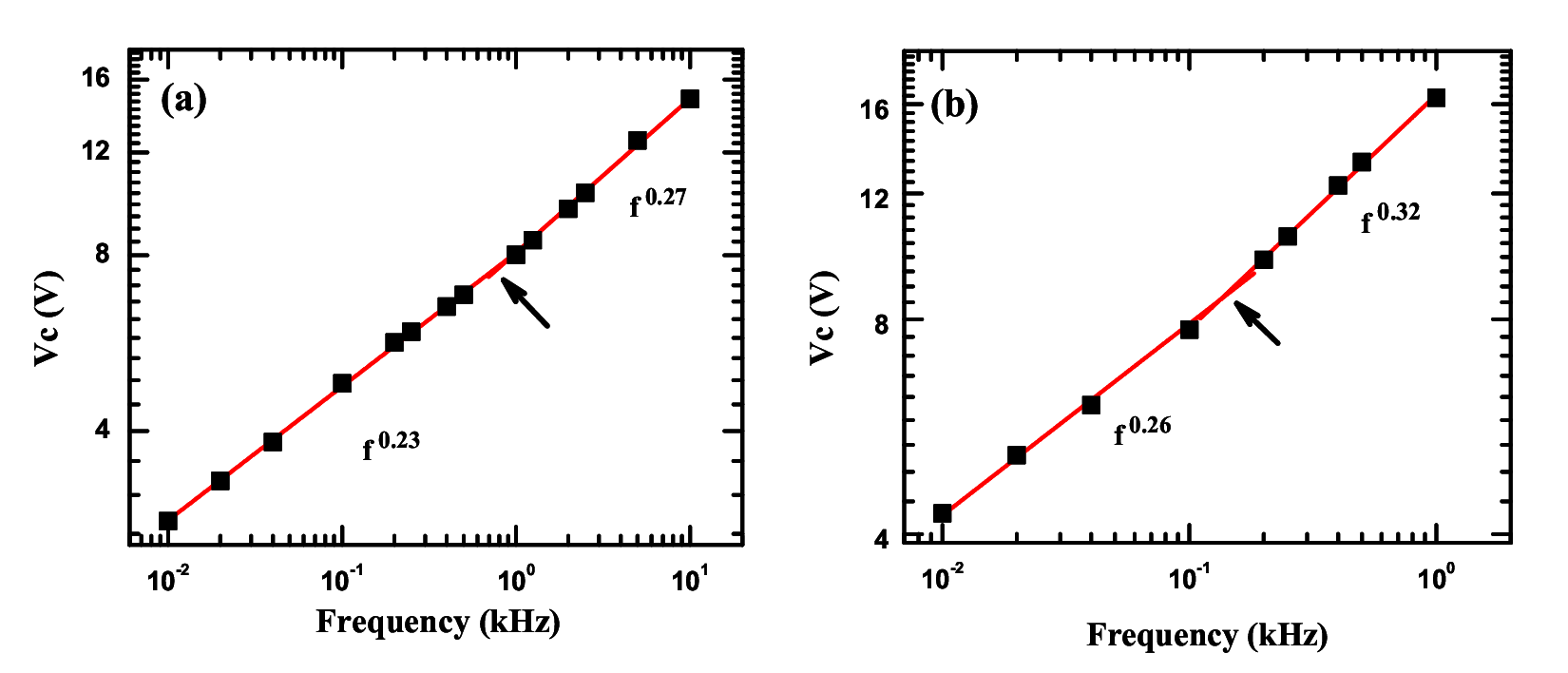}}
\caption{The coercive voltage $V_C$ versus frequency $f$ plots for (a) (111) and (b) (001) films; the data follow $V_C \propto f^{\$alpha}$ power law dependence; $\alpha$ changes at a specific frequency. }
\end{figure}     

\section{Experimental Details}
The La-doped Pb(Zr$_{0.4}$Ti$_{0.6}$)O$_3$ (PLZT) thin films and half-metallic bottom electrode layer of La$_{0.7}$Sr${0.3}$MnO$_3$ (LSMO) were grown on (001) and (111) oriented SrTiO$_3$ (STO) single-crystal substrates using a commercially available pulsed laser deposition (PLD; Neocera Pioneer 120 Advanced) system. These compounds have been chosen since earlier research \cite{Mukherjee,Rema} on these compounds shows that on doping the PZT with the donor dopant La$^{3+}$, the La atoms act as a charge compensator and reduce the mobile oxygen vacancies in the film which reduces the effect of leakage current. Lower doping helps to enhance the ferroelectric polarization. On the other hand, higher doping increases the defect density and pinning at these sites. PZT has also been extensively studied in the context of domain dynamics and polarization switching kinetics making it a suitable platform to study the transient negative capacitance. Prior to the film deposition, STO single crystal substrates were annealed inside the PLD chamber at 800 $^o$C temperature under an ambient oxygen pressure of 1000 mTorr for 2h. High-quality ceramic targets of (Pb$_{0.99}$La$_{0.1}$)(Zr$_{0.4}$Ti$_{0.6}$)O$_3$ (PLZT) and La$_{0.7}$Sr$_{0.3}$MnO$_3$ (LSMO) were sequentially ablated using a KrF excimer laser ($\lambda$ = 248 nm, frequency = 10 Hz, fluences = 3.5 J/cm$^2$) inside a deposition chamber equipped with a multitarget carousel allowing the in-situ deposition of multilayers with clean interfaces. The substrate-target distance has been kept constant at 5 cm throughout the deposition. The bottom LSMO layer (thickness $\sim$8-10 nm) was deposited at 800 $^o$C under 10 mTorr; this is followed by the deposition of PLZT layer (thickness $\sim$100 nm) at 550 $^o$C under 1000 mTorr. Open access to the bottom LSMO electrode was maintained throughout the process using shadow masks. Subsequently, top LSMO electrodes (200 $\mu$m diameter) were deposited using a shadow mask at 750 $^o$C under 10 mTorr. Following the deposition, the samples were cooled gradually to room temperature (duration of cooling was approximately 3h) under 1000 mTorr. The shadow mask that was used for electrode deposition was kept in close contact with the film surface in order to minimize the shadow effect. Two different films were used for the experiments with, respectively, [001] and [111] axes perpendicular to the surface of the film; hereinafter, they will be designated as (001) and (111) films. The films were characterized by x-ray diffraction - $\theta$-2$\theta$ scan and reciprocal space mapping (RSM). The X-ray photoelectron spectra (XPS) for the films were recorded at the PHI 5000 Versa Probe-II under a 3.7 $\times$ 10$^{-8}$ Pa ultra-high vacuum. A focused X-ray source of Al $K\alpha$ (1486.7 eV) was used to probe the core levels. The survey spectra were recorded by keeping the analyzer pass energy at 187.85 eV with a step size of 0.2 eV while for the core levels, the pass energy was set to 11.57 eV at a step size of 0.05 eV for reaching maximum resolution of 300 meV. 

For the measurement of the ferroelectric hysteresis loops, the bias voltage ($V$) was applied perpendicular to the film surface and therefore, the corresponding electric field $\textbf{E}\parallel[001]$ and $\textbf{E}\parallel[111]$. We employed standard triangular bipolar voltage pulse from a ferroelectric hysteresis loop tester (Precision LC-II of Radiant Inc.) for measuring the polarization ($P$) versus voltage ($V$) hysteresis loops. We have used a probe station to position the probes right onto the center of the top electrodes. We have repeated the measurements using different top electrodes and negligible variation in the ferroelectric hysteresis loop shape as well as in the magnitude of the saturation polarization could be noticed. 

\section{Computational Details}

The evolution of the ferroelectric domain dynamics was simulated using a two-dimensional time-dependent Ginzburg-Landau model, implemented on a square grid to solve the polarization field $P(x,y)$. The total free energy density $u$ of the ferroelectric system was formulated as:

\begin{equation}
u = aP^2 + bP^4 + cP^6 - EP + k(\nabla P)^2
\end{equation}

where $a$, $b$, and $c$ are Landau expansion coefficients representing ferroelectric anisotropy, $E$ is the externally applied electric field, and $k$ is the gradient energy coefficient accounting for domain wall energy. Only the out-of-plane polarization component was considered, assuming that the thin film was $c$-axis oriented and primarily underwent $180^\circ$ switching.

The polarization dynamics was governed by the Landau-Khalatnikov equation:

\begin{equation}
\frac{\partial P}{\partial t} = -\frac{1}{\rho} \frac{\delta F}{\delta P}
\end{equation}

where $\rho$ is a damping parameter related to internal loss and $F$ is the total free energy of the system. Spatial derivatives were computed using a finite-difference scheme on a uniform grid with fixed boundary conditions. Each cell $i$ was assigned local polarization $P_i$ and energy density $u_i$, with interaction terms involving nearest neighbors to model domain coupling.

The effective electric field $E_{eff}$ inside the ferroelectric was computed as:

\begin{equation}
E_{eff} = \frac{V_F}{t_F} - E_{bias}
\end{equation}

where $V_F$ is the voltage across the ferroelectric, $t_F$ is the film thickness, and $E_{bias}$ accounts for internal bias fields due to asymmetries at the electrode interfaces.

To capture the local inhomogeneities and nucleation behavior, spatial fluctuations in $a$, $b$, $c$, $q$, and $E_{bias}$ were introduced using Gaussian or uniform distributions. All the parameters were taken from the work of Hoffman $\textit{et al}$. \cite{Salahuddin}.

\section{Results and Discussion}
In Fig. 1, we show the $\theta$-2$\theta$ and RSM data for both the films. The structure is considered to be pseudocubic as the films grow epitaxially on cubic SrTiO$_3$ substrates. The insets of Figs. 1(a),(b) show the results of the detector scans about asymmetric (111) and (210) planes as well as about symmetric (002) and (111) planes for, respectively, the (001) and (111) films. Figures 1(c),(d), on the other hand, show the RSM about asymmetric (113) and symmetric (111) planes for, respectively, the (001) and (111) films. The lattice parameters were estimated to be $a$ = 4.00 ($\pm$0.007), $c$ = 4.09 ($\pm$0.002) [$c/a$ = 1.02] and $a$ = 4.08 ($\pm$0.002), $c$ = 4.09 ($\pm$0.003) [$c/a$ = 1.002] for the (001) and (111) films. The $c/a$ ratio is the `pseudotetragonality’ of the structure which measures the extent of distortion from the cubic structure \cite{Luo}. The $c/a$ ratio deviates from 1.00 in non-cubic structures. In the present case, while the $c/a$ ratio is close to 1.00 (actually 1.002) for the film with (111) plane orientation, it is higher (1.02) for the film with (001) plane orientation. It indicates that the film with (001) plane oriented parallel to the surface of the film is more distorted compared to the undistorted cubic structure. Smaller lattice mismatch between LSMO and STO induces larger pseudotetragonality in the (001) film while the converse is true for the (111) film resulting in a more relaxed state with lower pseudotetragonality. The thickness of PLZT for both films (001) and (111) is $\sim$100 nm and the roughness determined using the atomic force microscopy topography images, recorded from different positions of the films, varies within 1.50-7.00 nm. The thickness of film we have used in our study has widely been reported for observing strain relaxation and robust ferroelectricity with minimum leakage current density. Since the roughness of the film surface is less than 10\% of the film thickness, the surface is relatively smooth and the spatial uniformity of applied electric field is maintained which is essential for observing consistent results in the case of the transient negative capacitance measurements. The XPS data for the La$^{3+}$, Pb$^{2+}$, Ti$^{4+}$, and O$^{2-}$ ions are shown in the supplementary materials document. Fitting of the peaks by Peak 41 software reveals that the oxygen stoichiometry in the films is different. While in (001) film, the oxygen vacancy concentration is $\sim$11\%, it is $\sim$7\% in the (111) film. This accounts for a bit higher leakage in the (001) film.  

Figures 2a and 2c, respectively, show the $\textit{major}$ $P-V$ hysteresis loops traced over different timescale (across 200 $\mu$s-100 ms) for the (111) and (001) films while Figs. 2b and 2d show the corresponding timescale dependence of the coercive voltage $V_C$; $V_C$ is higher at a smaller timescale (i.e., at a higher frequency). The $V_C$ versus frequency ($f$) data are found to follow a power-law dependence $V_C$ $\propto$ $f^{\alpha}$ (Fig. 3). It is important to point out here that even though the difference in deposition temperature of top and bottom electrodes could introduce asymmetric interface barriers or dead layers, its influence on the ferroelectric properties such as vertical asymmetry \cite{Borisov} in the ferroelectric hysteresis loop, imprint effect, change in ferroelectric polarization and coercive voltage over a number of cycle of measurements etc. turns out to be quite small. We did not observe vertical asymmetry in the hysteresis loop. The imprint effect is also quite small. Similar effect has been observed by others as well. Moreover, all the results are consistent across a number of cycle of measurements. This could be because the films were annealed following deposition using appropriate partial pressure of oxygen within the deposition chamber which helped in reducing the oxygen non-stoichiometry and also the asymmetry in the film/electrode interface barrier composition \cite{Mak}. Interestingly, for both the cases of $V_C - f$ patterns, there is a crossover in the power-law dependence and $\alpha$ varies from 0.23 to 0.27 and from 0.26 to 0.32 for (111) and (001) films, respectively. It indicates a crossover in the domain switching mechanism (discussed later) across the timescale range covered here. The magnitude of the $V_C$ is large (of the order of 1000 kV/cm). However, such a large $V_C$ has earlier been observed \cite{Borderon} in PZT system containing, primarily, c-domains because of higher domain pinning energy. Small gap observed in the $P-V$ hysteresis loops could originate from a little bit of leakage \cite{Meyer}. The leakage current versus voltage characteristics were measured for both the samples (supplementary materials); it turns out to be three orders of magnitude smaller than the polarization current.  The $P-V$ loop shape conforms to the expected shape in most of the cases except when traced at a faster rate in the case of the (001) film. It cannot be due to the influence of leakage since at a shorter timescale the influence of leakage decreases \cite{Meyer}. The `gap' in the $P-V$ loop decreases progressively with the decrease in timescale of the measurement (supplementary materials). Also, using two sets of preset pulses without any intermediate delay, it is possible to show that the `gap' in the $P-V$ loop could be nearly eliminated (supplementary materials). Importantly, determination of the capacitance-voltage characteristics from the differentiation of the $P-V$ loops shows clearly negligible influence of the gap on the specific capacitance. In addition, measurement of the hysteresis loop has been carried out using a specially designed bias voltage pulse structure which eliminates the lossy component effectively (supplementary materials). The results of all these measurements indicate that the distortion in the loop shape cannot originate from leakage. The unusual loop shape observed under shorter voltage pulse (less than 1 ms) could possibly result from change in the domain switching mechanism and the corresponding thermodynamics. Distorted $major$ $P-V$ loops at a higher voltage and shorter timescale have been observed by others as well \cite{Hu}. 

\begin{figure*}[ht!]
\begin{center}
   \subfigure[]{\includegraphics[scale=0.45]{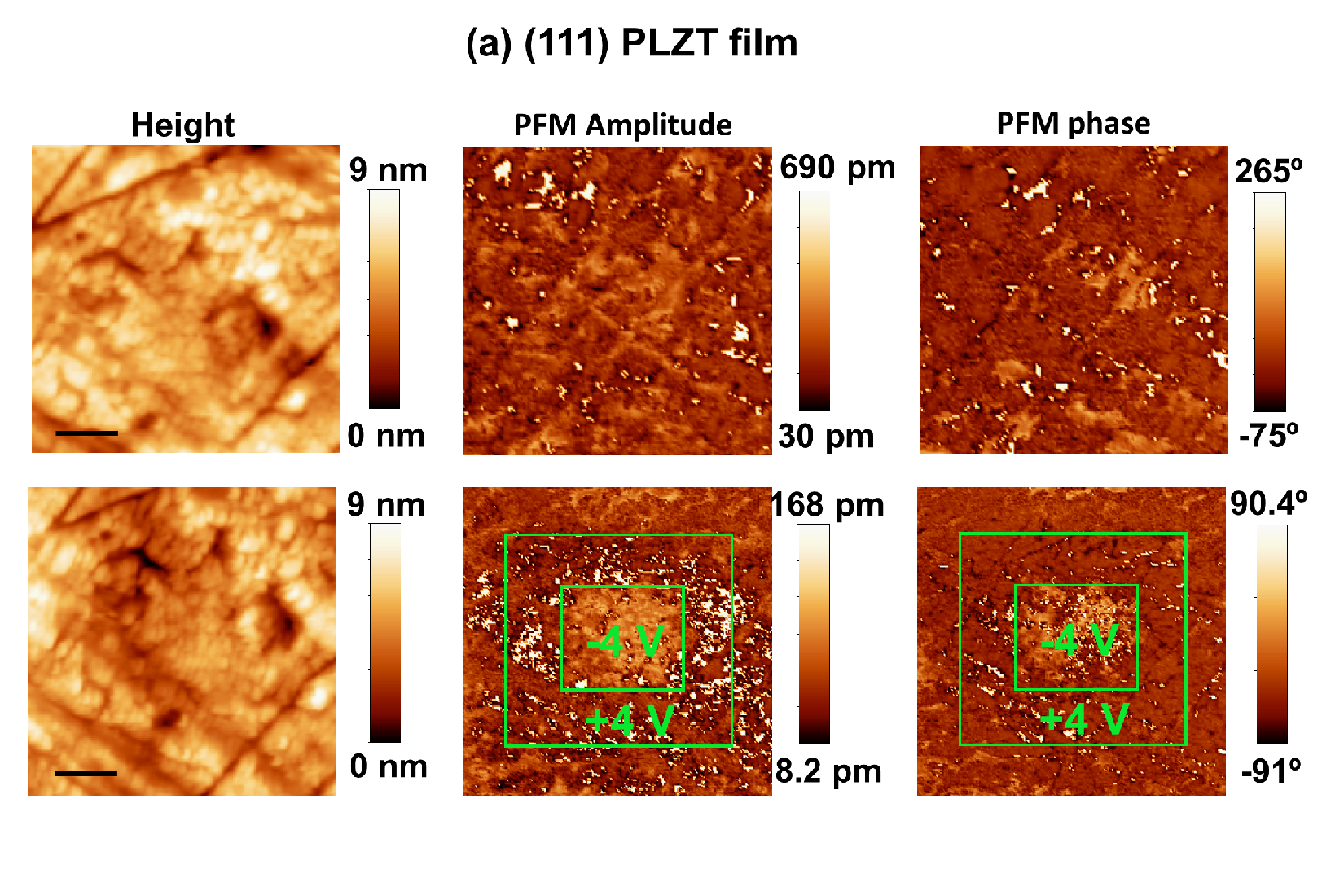}} 
   \subfigure[]{\includegraphics[scale=0.45]{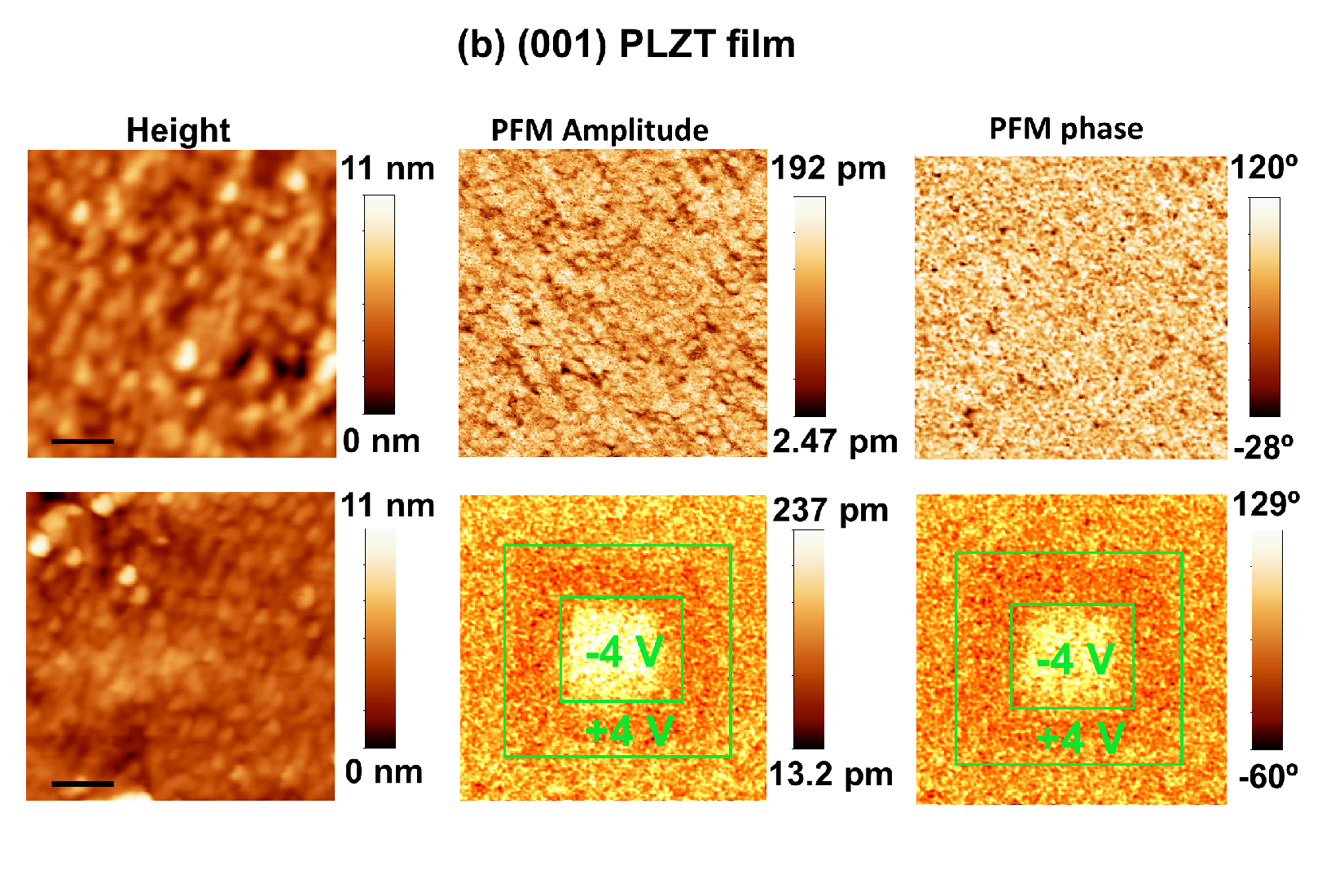}}
\end{center} 
\caption{(a) The height, PFM amplitude, and PFM phase images of (a) (111) and (b) (001) PLZT film. The upper panel PFM images are collected under zero bias tip voltage, whereas the lower panel images are collected under $\pm$4 V tip bias. Box-in-box pattern in the lower panel images are created by two steps electric poling with -4 V and +4 V tip bias. Scale bar is 1 $\mu$m.}
\end{figure*}

Piezoresponse force microscopy (PFM) and switching-spectroscopy PFM (SS-PFM) were carried out to examine the nanoscale ferroelectricity in the samples. The PFM images (upper panels of Fig. 4) exhibit a clear ferroelectric response. Polarization switching was also observed (lower panels of Fig. 4) in the regions poled using a two-step biasing process with +4 V and -4 V applied through the conducting tip. The SS-PFM hysteresis loops are shown in Fig. 5 and they conform to the expected behavior signifying the presence of robust nanoscale ferroelectricity. To support this, we have quantified the phase response in the switched and unswitched regions. The average phase difference was found to be approximately 90.4$^o$-(-91$^o$) for the (111)-oriented and 129$^o$-(-60$^o$) for the (001)-oriented films, indicating near-complete ($\sim$180$^o$) ferroelectric switching. Further, line scans have been taken across the regions of phase-contrast images poled under +4 and -4 V. The results show reversal of domains (supplementary materials). There is a discrepancy in the coercive voltage ($V_C$) determined from SS-PFM and macroscopic electrical measurements. It can be attributed to the differences in spatial and temporal resolution between the two techniques. SS-PFM is a local probe method that measures switching behavior over a small volume beneath the tip, whereas the electrical measurements reflect the global behavior of the entire sample. Additionally, SS-PFM is performed with a slower, quasi-static voltage sweep, which can influence the apparent switching characteristics. The timescale dependence of $V_C$ has been observed in electrical measurements (Figs. 2b and 2d). These factors are likely to be contributing to the observed differences in $V_C$.

\begin{figure*}[ht!]
\centering
{\includegraphics[scale=0.35]{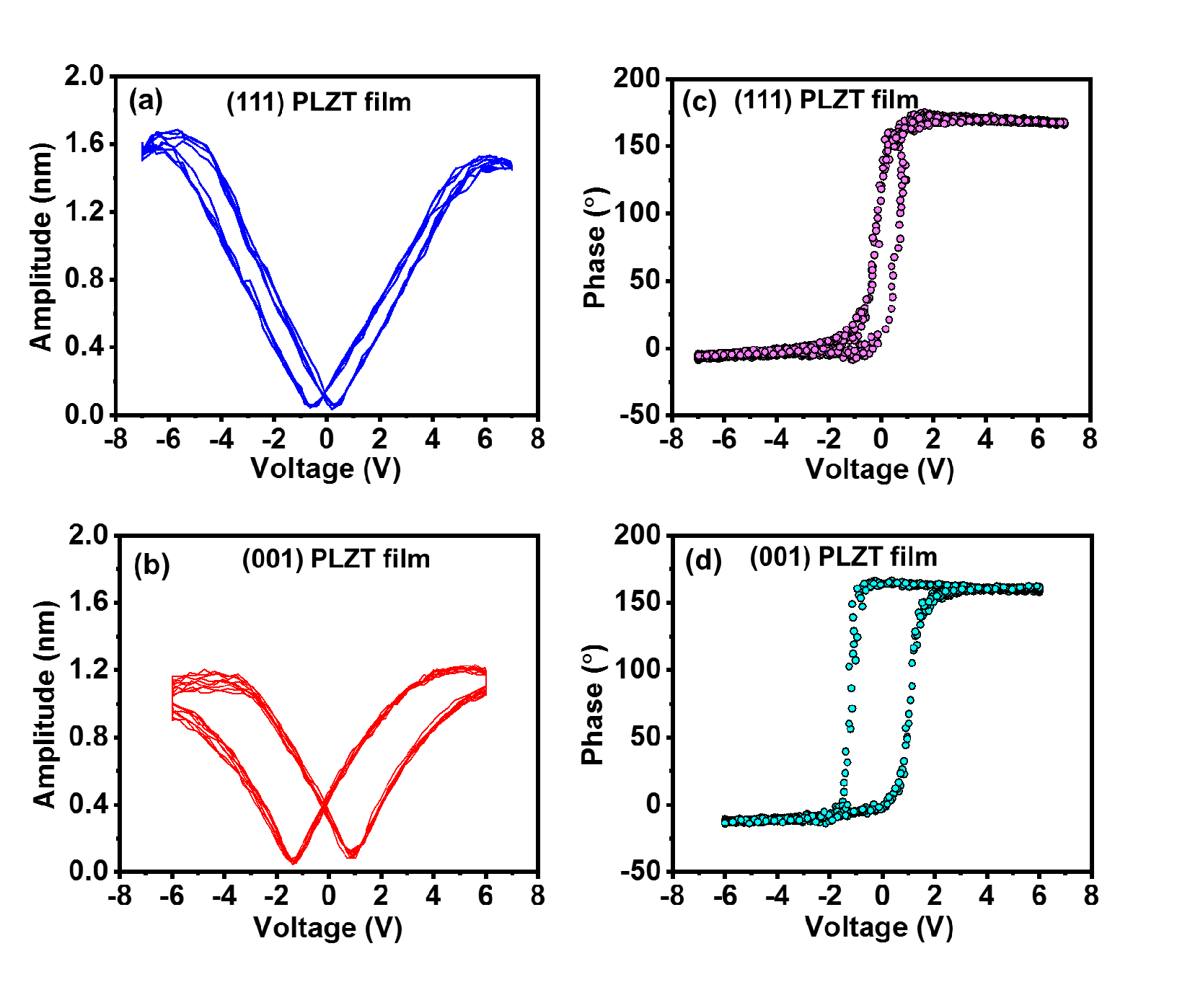}}
\caption{The PFM switching-spectroscopy (PF-SS) on an arbitrary point of both the films. (a)-(b) Local ferroelectric butterfly-shaped amplitude loops and (c)-(d) local hysteresis phase loops as a function of the applied voltage for (111) and (001) films, respectively. }
\end{figure*}

\begin{figure*}[ht!]
\centering
{\includegraphics[scale=0.70]{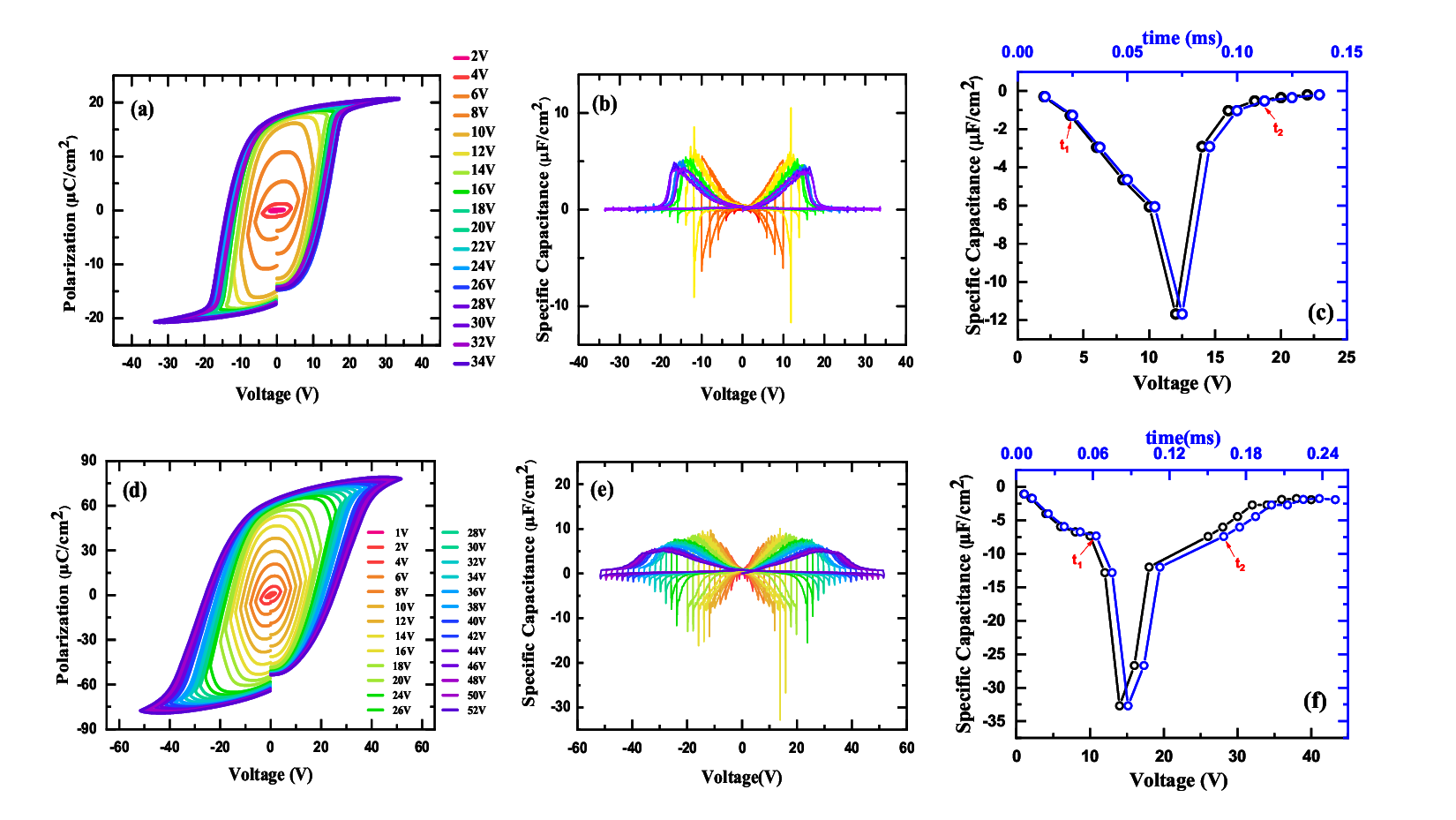}}
\caption{(a),(b),(c) The minor to major $P-V$ loops (recorded at 0.5 ms time scale), $C-V$ plots (obtained from differentiating the $P-V$ data), and negative $C$ versus coercive voltage and scan time patterns for the (111) film; (d),(e),(f) show the corresponding plots for the (001) film; minor to major $P-V$ loops were traced at 1 ms time scale. }
\end{figure*}

The $\textit{minor}$ to $\textit{major}$ $P-V$ hysteresis loops have been traced over 2-50 V and across 200 $\mu$s - 100 ms. The minor loops do not signify ferroelectric domain switching. Therefore, the negative capacitive state cannot arise so long as the ferroelectric domains do not switch and cross the energy barrier. The negative capacitance becomes significant and meaningful only when the hysteresis loops cross over to the regime of major from the one of minor. It is important to trace the hysteresis loops in such a way so that the evolution of the major loops from the minor ones could be clearly mapped. This is because the negative capacitance becomes significant right at the crossover point between the minor and major loop regimes. In order to capture this crossover point, the entire set of minor and major loops have been traced keeping the timescale of the bias voltage pulse constant. This helps in determining the negative capacitance for a specific coercive voltage corresponding to the specific timescale. In Figs. 6a and 6d, we show the representative sets of $P-V$ loops recorded over 0.5 and 1.0 ms for (111) and (001) films. The supplemental materials document (supplementary materials) contains additional set of loops recorded at different time scale. Differentiating the $P-V$ loops we obtain the specific capacitance $C$ versus $V$ patterns (Figs. 6b and 6e). It appears that the specific capacitance ($C$) becomes $\textit{negative}$ around the $P = 0$ state of the loops and it reaches maximum ($C_m$) (Figs. 6c and 6e) at the coercive voltage $V_C$ \cite{Pintilie} corresponding to the timescale of the measurement (Fig. 2). Using the voltage ($V$) - timescale ($t$) profile, $V$ has been converted to $t$ ($t$ = $V/4fV_{max}$, $f$ = frequency) and in Figs. 6c and 6e, negative $C$ versus $t$ plots are also shown. The range marked in the plots ($t_1$ and $t_2$) yields the domain switching time scale ($t_2 - t_1$); it turns to be $\sim$0.1 ms.   

\begin{figure*}[ht!]
\centering
{\includegraphics[scale=0.50]{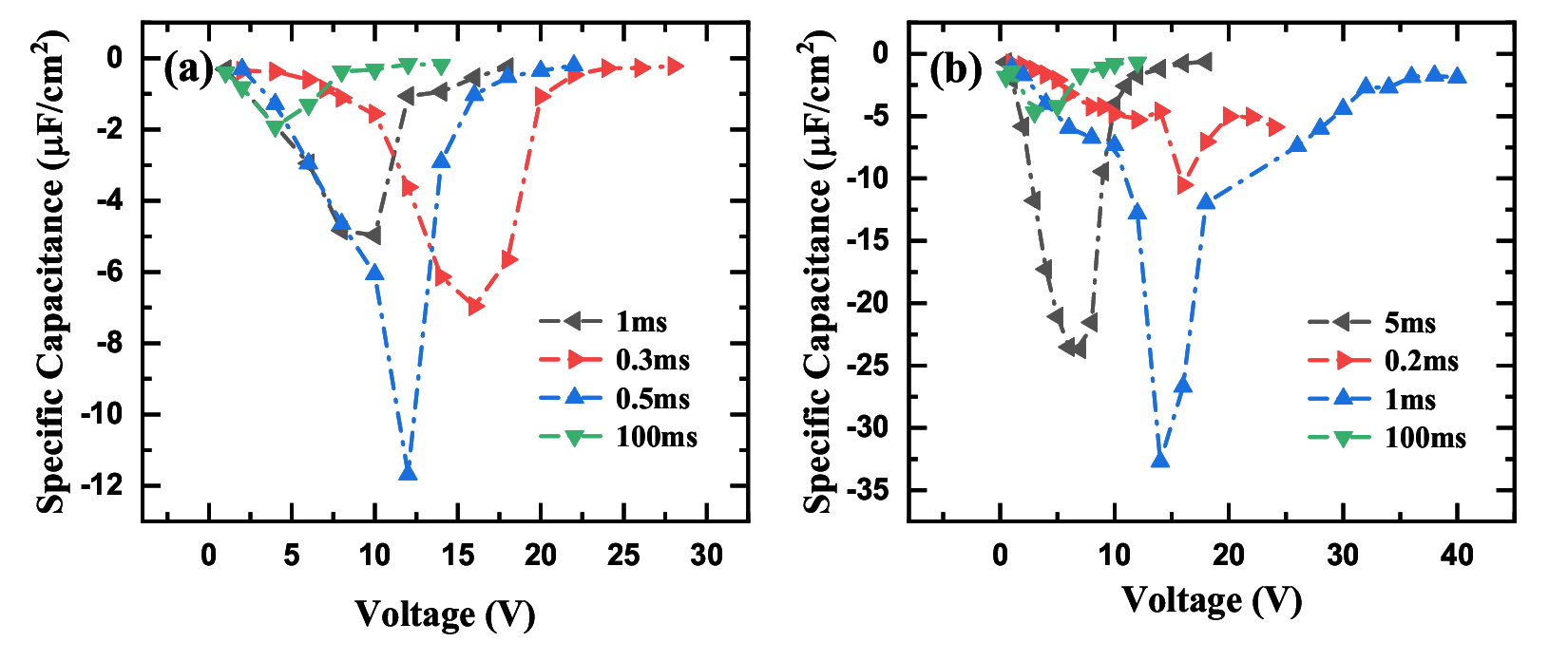}}
\caption{The specific capacitance versus coercive voltage patterns obtained from the $P-V$ loops recorded at different time scale for (a) (111) and (b) (001) films; the negative capacitance maximizes when the loops are traced over a specific time scale - 0.5 ms for (111) film and 1 ms for (001) film. }
\end{figure*}

We derived the timescale-dependent $C-V$ patterns from the $P-V$ loops measured across different timescale. Representative plots are shown in Fig. 7. Interestingly, we observed (i) remarkable bias voltage amplitude and timescale dependence of specific negative capacitance and (ii) maximization of the specific negative capacitance ($C_{m_c}$) at a specific bias voltage amplitude and timescale. For instance, the specific negative capacitance maximizes when the measurements were carried out at 0.5 and 1.0 ms for (111) and (001) films, respectively. The $P-V$ loops measured at a faster or slower rate yield comparatively smaller specific negative capacitance. \textit{This is one of the key results of this work}. In this context, it is important to mention that the ferroelectric loop tester (Precision LC-II model of Radiant Technologies Inc.) uses virtual ground-charge integration where the input point is actively held at the zero potential. Here the parasitic capacitance is considered to be between the virtual ground input and earth ground. Since there is no voltage across the parasitic capacitor, it does not introduce any contribution to the signal. Therefore, the timescale of response (polarization or capacitance) corresponds to the intrinsic timescale of polarization switching and not that of the response of circuit elements or parasitic components. It is also worth mentioning that the coercive voltage and timescale corresponding to the maximum transient capacitance mark the crossover point in the $V_C-f$ plot, i.e., the point at which change in the exponent $\alpha$ of the power-law dependence of $V_C-f$ data takes place. We further examined the issue of maximum negative capacitance at a specific bias voltage amplitude and timescale by measuring the differential voltage amplification \cite{Khan} over different timescale. In Fig. 8, we show the representative $V_d$, $V_f$,and $V_s$ versus $t$ plots; $V_s$, $V_f$, and $V_d$, respectively, are the source voltage and the voltage drops across the PLZT ferroelectric capacitor and a dielectric capacitor of known capacitance connected in series. The amplification ($A_v$ = $\Delta V_d/\Delta V_s>$1.0; $\Delta V_s$ is the change in source voltage over a certain time scale) of the $V_d$ implies drop in the $V_f$; this, in turn, signifies transfer of energy from the ferroelectric to the dielectric capacitor. Expectedly, in this case too, we noticed maximization of the $A_v$ (Fig. 8) when the measurements were carried out across specific time scales - for instance, across 0.5 ms for (111) film and 1.0 ms for (001) film. It implies maximization of $A_v$ due to maximization of the negative capacitance. The plot of charge $Q$ versus $V_f$ for the segment across which the voltage amplification has been observed is shown in the supplemental materials document. Clearly, the $Q-V_f$ plot exhibits negative curvature. Both of these set of results (Figs. 7 and 8) signify bias-voltage-pulse dependent specific negative capacitance in PLZT ferroelectric capacitor and its maximization at a characteristic pulse profile (amplitude and timescale). Of course, the extent of amplification $A_v$ [varying within $\sim$1.01-1.03 (supplementary materials)], though reproducible, is smaller in the case of the (001) film. The difference in (i) the magnitude of the maximum specific negative capacitance, (ii) the bias voltage and timescale at which the maximum negative capacitance is observed, and (iii) the magnitude of differential voltage amplification ($A_v$) observed in (111) and (001) films could possibly be due to difference in the (i) angle between the direction of application of bias voltage and polarization axis, (ii) structural distortion [(001) film is more distorted], and (iii) the defect concentration and its contribution to the domain switching. The smaller $A_v$ in the case of the (001) film could originate from intrinsic features as well as non-ideal voltage distribution and influence of circuit parameters even though attempts were made to keep those influences small.   

\begin{figure*}[ht!]
\centering
{\includegraphics[scale=1.60]{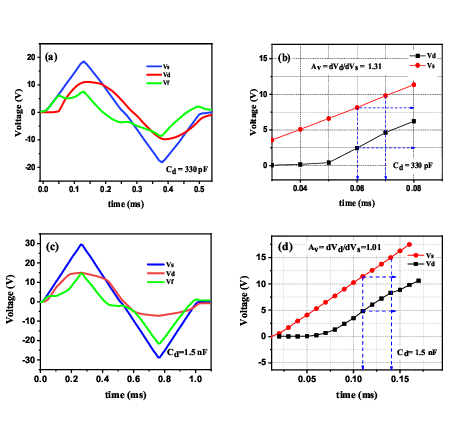}}
\caption{(a),(b) The source voltage ($V_S$) and the voltage drop across the ferroelectric and dielectric capacitors ($V_F$, $V_d$) versus $time$ plots for the (111) film; (c),(d) corresponding plots for the (001) film; (b) and (d) show clearly the extent of differential voltage amplification in both the cases. }
\end{figure*}  

\begin{figure*}[ht!]
\centering
{\includegraphics[scale=0.45]{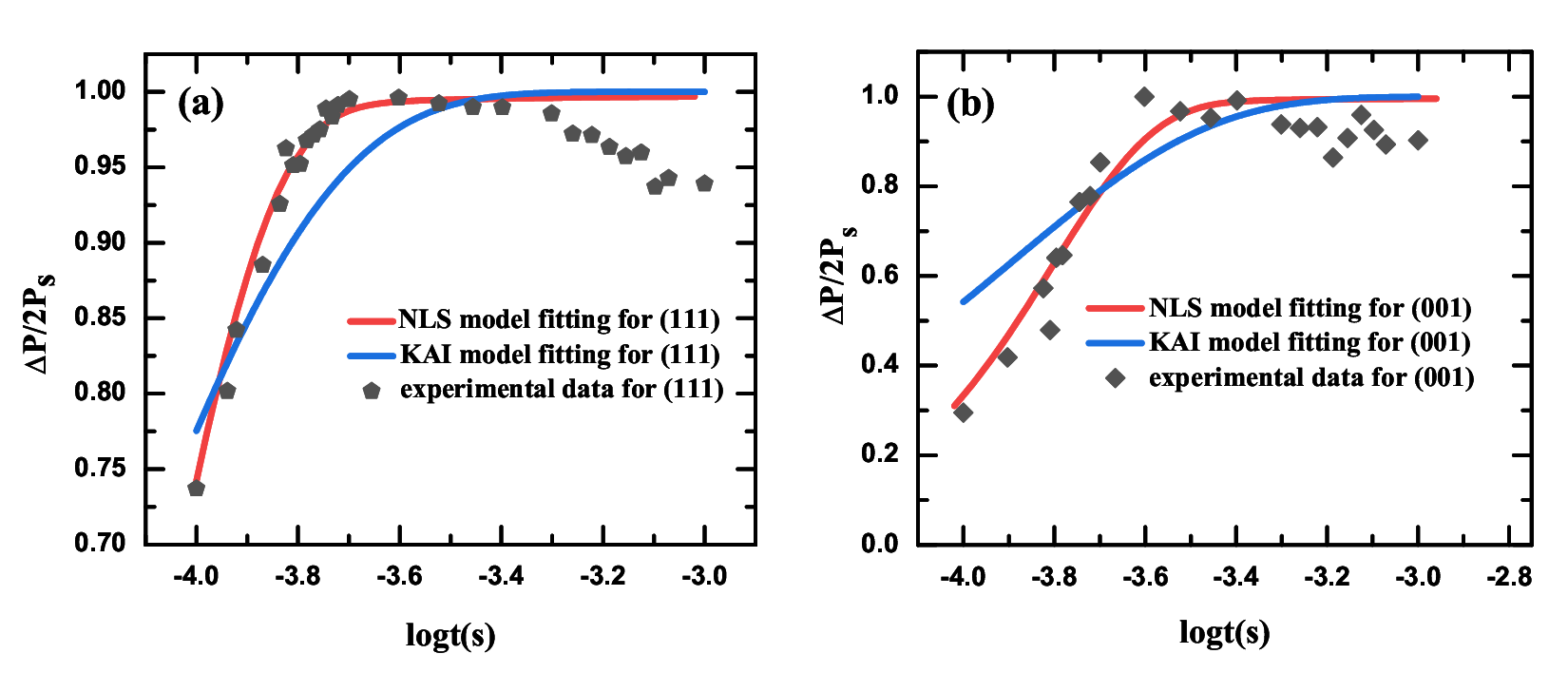}}
\caption{The polarization switching versus time scale plot and their fitting by KAI and NLS models for (a) (111) and (b) (001) films. }
\end{figure*}
 
In order to trace the origin of maximization of negative capacitance at a specific bias voltage and timescale, we extracted the characteristic intrinsic domain switching kinetics in a separate experiment. The switched polarization ($\Delta P$) was noted by sending a rectangular positive-up-negative-down (PUND) pulse profile of different time scale (100 $\mu$s-5 ms). Figure 9 shows the normalized $\Delta P$ versus $t$ plots recorded across 100 $\mu$s-5 ms at an applied $V$ $\geq$ $V_C$. The patterns are consistent with those observed by others \cite{Noh}. However, it could not be fitted by the well-known Kolmogorov-Avrami-Ishibashi (KAI) model of homogeneous nucleation and growth, $\Delta P(t) = 2P_S[1 - exp\{-(t/t_0)^n\}]$. Instead, as pointed out in Ref. 20, an inhomogeneous switching model - nucleation limited switching (NLS) - with a Lorentzian distribution of the switching time appears to be more suitable in these cases too. In this model, the switched polarization ($\Delta P$) is given by\\ 

\begin{equation}
\Delta P(t) = 2P_s\int_{-\infty}^{\infty}[1 - exp\{-(t/t_0)^n\}]F(logt_0)d(logt_0) 
\end{equation}

where $F(logt_0)$ describes the Lorentz distribution 

\begin{equation}
F(logt_0) = \frac{A}{\pi}[\frac{w}{(logt_0 - logt_1)^2 + w^2}]
\end{equation}

The mean switching timescale of the Lorentz distribution $t_p$ [$\sim$0.4 and $\sim$0.8 ms for (111) and (001) films, respectively], obtained from the fitting, turns out to be close to the timescale at which the negative capacitance and differential voltage amplification maximize. This close correlation is indeed remarkable. It points out that when the bias voltage pulse profile induces ferroelectric domain switching via nucleation and growth of reverse domains \cite{Li}, the domain wall density maximizes during switching and this, in turn, results in maximum negative capacitance. Domain switching kinetics, measured at different timescale and under different bias voltages (supplementary materials), results in faster or imcomplete switching, as observed by others \cite{Noh}.   

Very interestingly, the shape of the resulting $P-V$ hysteresis loop too, turns out to be dependent on the specific transient negative capacitance. The radius of curvature ($r$) of the $P-V$ loop ($r \propto [\partial^2P/\partial V^2]^{-1}$) at $V_C$ and its inverse $|\partial C/\partial V|_{V_C}$ (i.e., the rate of change of capacitance with voltage at the coercive voltage) exhibit a systematic scaling with the negative capacitance - larger the negative capacitance, smaller is the $r$ and larger is the rate of change of $C$ at $V_C$ (Fig. 10). It is important to mention, in this context, that the (001) film, possibly because of higher pseudotetragonality, exhibits domain switching over a broader time scale range [in comparison to the (111) film]. This is reflected in the time scale range - $\sim$1.0-5.0 ms - at which the specific negative capacitance, differential voltage amplification (supplementray materials), and $|\partial C/\partial V|_{V_C}$ maximize while $r$ of the $P-V$ loop minimizes.

\begin{figure*}[ht!]
\centering
{\includegraphics[scale=1.55]{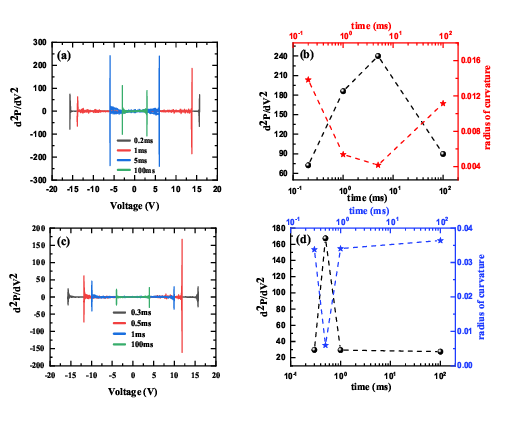}}
\caption{(top panels) (left) The double derivative of the P-V loops recorded at different time scale for the (001) film; (right) variation of the double derivative and its inverse at VC with the time scale. (bottom panels) (left) The double derivative of the P-V loops recorded at different time scale for the (111) film; (right) variation of the double derivative and its inverse at VC with the time scale.   }
\end{figure*}

\begin{figure}[ht!]
\centering
{\includegraphics[scale=0.30]{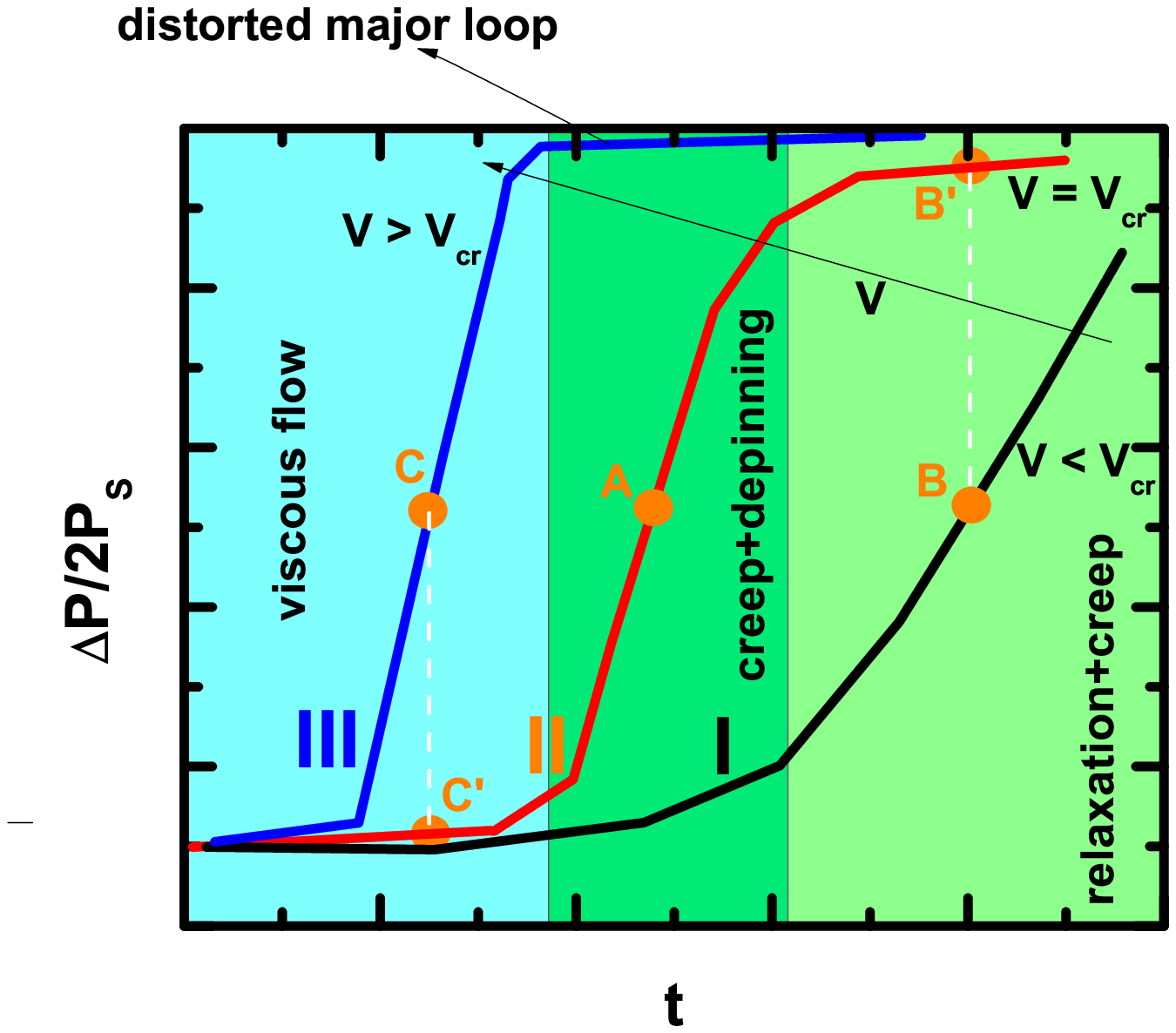}}
\caption{Schematic of different domain switching pathways - I, II, and III. The ``switching" state of pathways I and III - points B and C - corresponds to the ``switched" (point B') and ``unswitched" (point C') states of thermodynamically preferred pathway II. }
\end{figure}

\begin{figure*}[ht!]
\begin{center}
   \subfigure[]{\includegraphics[scale=2.65]{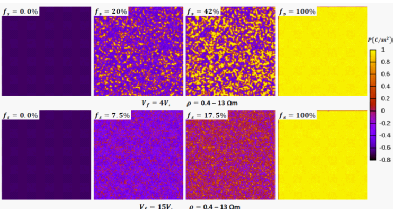}} 
   \subfigure[]{\includegraphics[scale=2.65]{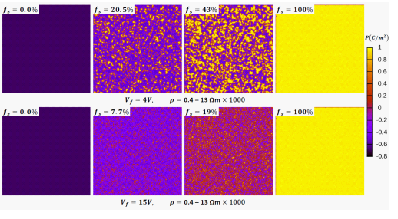}}
\end{center} 
\caption{Temporal evolution showing the spatial polarization distribution in the ferroelectric during switching for (a) $\rho = 0.4$–13 $\Omega\cdot$m and (b) $\rho = 0.4$–13 $\Omega\cdot$m $\times$ 1000, respectively.}
\end{figure*}

\begin{figure*}[ht!]
\begin{center}
   \subfigure[]{\includegraphics[scale=0.65]{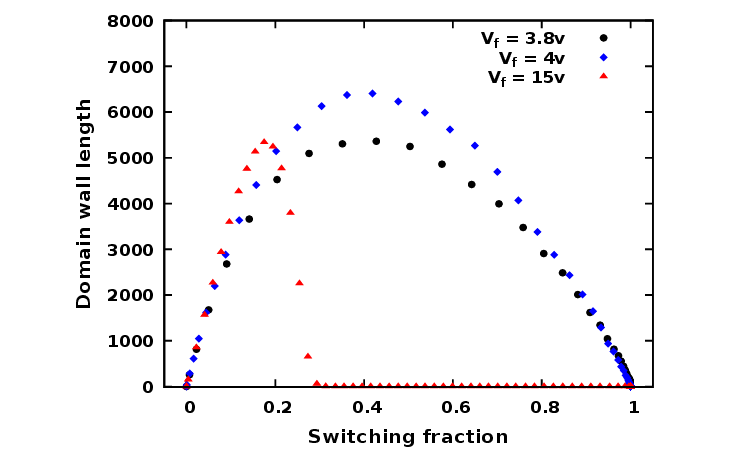}} 
   \subfigure[]{\includegraphics[scale=0.65]{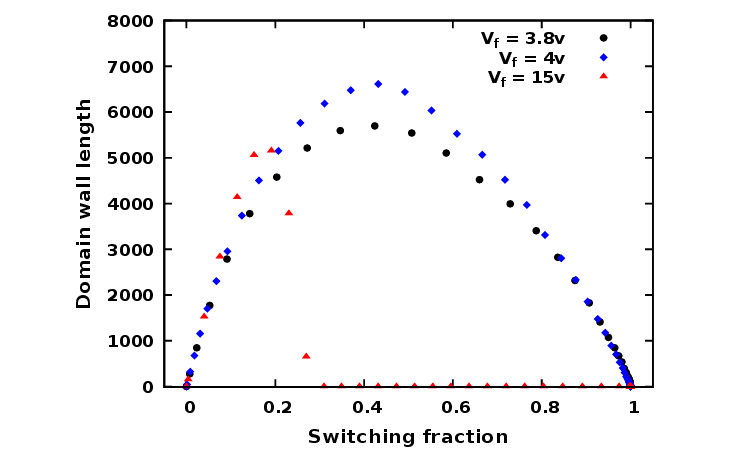}}
\end{center} 
\caption{Domain wall length as a function of switching fraction for  (a) $\rho = 0.4$–13 $\Omega\cdot$m and (b) $\rho =0.4$–13 $\Omega\cdot$m $\times$ 1000, respectively.}
\end{figure*}

To explain this remarkable dependence of transient negative capacitance on bias voltage pulse profile and correlation with domain switching kinetics and shape of the $P-V$ hysteresis loop, we point out that the switching of ferroelectric domains involves both nucleation and growth of reverse domains and free energy landscape (i.e., energy barrier between the switched states) \cite{Li}. While the nucleation is governed by defect centers, the free energy landscape is determined by the structural aspects and types of domains. Variation of bias voltage results in variation of the switching timescale as well as the pathway of switching. For instance, in the PZT systems, earlier work \cite{Gao} has shown that the presence of 90$^o$ ferroelastic domain walls could induce complicated path of domain switching which requires higher bias voltage. It has also been pointed out \cite{Xu} that application of different bias voltage pulse profile results in different domain configuration (in other words, different ``switched" states). Intrinsically, the domain switching is associated with crystallographic transition and different switching pathways are associated with different energy barrier \cite{Beckman}. The energy barrier also depends on the strain of the sample \cite{Beckman}. For other ferroelectrics too, tracking of structural transition during domain switching yields the sequence of crystallographic structures emerged and transitions during the intermediate state of ``switching"; the energy barrier for such transitions has also been determined \cite{Lee,Bdikin,Zhang,Fennie,Zhao,Britson,Franzbach,Heron}. All these results point out that (i) by varying the bias voltage amplitude and timescale it is possible to induce different domain switching pathways and (ii) since each pathway is associated with a specific energy barrier, the one with minimum energy barrier is thermodynamically favorable. The variation of specific negative capacitance as well as shape of the $P-V$ hysteresis loop with bias voltage pulse amplitude and timescale could, therefore, possibly due to variation in the domain switching pathways.     

In order to elucidate this point further, we mention that the intrinsic energy landscape of domain switching is modified because of extrinsic effects such as defect centers which act as nucleation as well as pinning centers of domain growth and switching \cite{Sear}. Earlier work on domain switching has shown that switching involves nucleation of reverse domains and their growth during which the domain wall density enhances \cite{Pan}. A recent work \cite{Li} has pointed out that both the nucleation and growth as well as intrinsic energy landscape associated with structural transition influence the domain switching kinetics. These two mechanisms are not mutually exclusive. Of course, it has been shown \cite{Li} that nucleation influences the switching kinetics at the initial stage of switching while influence of free energy landscape is observed at a later stage. Depending on the type of domain walls (e.g, 90$^o$ and 180$^o$ in the present case) as well as lattice strain and defect concentration of the samples (which act as domain pinning centers), the nucleation and growth processes assume specific (i.e., intrinsic) timescales. We showed above that inhomogeneous NLS mechanism explains the switching kinetics in the present case (Fig. 9) where the bias voltage and timescale exhibit Lorentzian distribution. The mean voltage $V_m$ and $t_m$ represent the parameters for which the domain-wall length during ``swithching" and hence the transient negative capacitance maximizes. Variation of bias voltage pulse profile over a wide range (from lower to higher voltage and shorter to longer timescale) triggers ``switching" with varying proportion of domains in ``unswitched", ``switching", and ``switched" states. For three cases - where the bias voltage ($V$) and timescale ($t$) are (i) $<$ $V_m$, $>$ $t_m$ or (ii) $>$ $V_m$, $<$ $t_m$ or (iii) $\approx$ $V_m$, $\approx$ $t_m$ - the volume fraction of domains in (a) ``unswitched", (b) ``switching", and (c) ``switched" states are (a) $x_a$, $x_b$, $x_c$, (b) $y_a$, $y_b$, $y_c$, and (c) $z_a$, $z_b$, $z_c$ ($x_i +y_i +z_i$ = 1.0), respectively. From the Lorentzian distribution pattern of $V$ and $t$, it is possible to observe (albeit qualititavely) that $x_c$ $<$ $x_a$ $<$ $x_b$, $z_c$ $<$ $z_b$ $<$ $z_c$; therefore, $y_c$ $>$ $y_a$ and $y_b$. It points out that indeed the volume fraction of the domains in ``switching" states enhances only when the bias voltage pulse profile (amplitude and timescale) becomes comparable with the $V_m$ and $t_m$. It can be illustrated also by a schematic of the domain switching kinetics and the regimes of different switching mechanisms in the $\Delta P/2P_S$ versus $t$ plane (Fig. 11). Bias voltage pulses of different amplitude and timescale could induce different switching pathways - either I, II or III. A point in the ``switching" regime of thermodynamically preferred pathway (point A on pathway II) corresponds to the points B and C on the pathways I and III for $V$ $<$ $V_m$, $t$ $>$ $t_m$ and $V$ $>$ $V_m$, $t$ $<$ $t_m$ respectively. In effect, they signify ``switched" (point B') and ``unswitched" (point C') states, respectively, with respect to the pathway II. Therefore, a sizable fraction of the domains would energetically prefer to be in ``unswitched" and ``switched" states if the pathways I or III are followed. This is quite analogous to the glass transition where rapid solidification or quenching of liquid through the melting point results in kinetic arrest of liquidus or amorphous phase and does not allow nucleation and growth of crystalline phase \cite{Biroli,Debenedetti}. Obviously, in those states the domain wall density would be smaller than what one observes in the ``switching" regime of pathway II. In fact, the thermodynamically preferred switching pathway yields maximum number of nucleation sites (at the pinning sites) by lowering the activation energy of nucleation \cite{Yoon}. This optimum domain configuration yields the maximum specific negative capacitance. From both theroretical and detailed experimental work, it has earlier been shown that domain walls induce negative capacitance \cite{Ramesh,Salahuddin}. 

To investigate the influence of the applied voltage and damping parameter (i.e., defect density) on ferroelectric switching behavior quantitatively, we performed time-dependent Ginzburg-Landau simulations of domain evolution under two different applied voltages ($V_f = 4$V and $15$V), for two damping regimes: (i) $\rho = 0.4$–13 $\Omega\cdot$m and (ii) $\rho = 0.4$–13 $\Omega\cdot$m $\times$ 1000. We examine (i) how domain-wall length during switching depends on the bias voltage amplitude ($V_f$), (ii) how the timescale of switching too depends on the $V_f$, and finally (iii) how defect density (i.e., damping parameter) changes the timescale of switching for identical $V_F$, i.e., how the domain wall motion is influenced by the defect density. In the Figs. 12(a) and 12(b) the corresponding domain microstructure evolution snapshots  (50 $\mu m$ $\times$ 50 $\mu m$) are shown for the two damping regimes, respectively. At an ``optimum" voltage ($V_f = 4$V), the switching proceeds through a classical nucleation and growth mechanism. As seen in both the damping cases, isolated reversed domains emerge (at low switching fraction $f_s$) and expand laterally with time until full reversal ($f_s = 1$) is achieved. Domain coalescence occurs gradually, and the transformation exhibits strong spatial inhomogeneity, especially at lower damping where the domain wall mobility is higher. 

In contrast, under higher voltage ($V_f = 15$V), the system undergoes a more abrupt transition, where switching occurs through a collective transformation, passing through largely ``unswitched" intermediate states with fewer distinct nucleation centers. Particularly in the low-damping regime [Fig. 12(a)], the polarization reversal is more spatially uniform and rapid, suggesting the field is sufficiently strong to suppress nucleation-limited kinetics and induce direct switching of large regions. In the case of low damping parameter, the switching completes within a timescale of approximately 0.1–4.0 $\mu$s. However, under high damping parameter, this timescale extends significantly to the range of 100 $\mu$s–4.0 ms.

To quantify these differences, we analyzed the total domain-wall length as a function of switching fraction $f_s$ [Fig. 13 (a) and (b)]. In all the cases, domain wall length initially increases as reversed domains nucleate and grow, peaks when domain-wall density is highest, and then declines as domains merge. However, the peak domain-wall length occurs at markedly different switching fractions depending on the applied voltage. For $V_f = 4$ V, the domain wall length maximizes when the switched state $f_s$ $\approx$ 0.42, reflecting a progressive and distributed nucleation process. In contrast, for $V_f = 15$ V, the maximum in domain wall length is smaller and  shifts significantly earlier, occurring at $f_s$ $\approx$ 0.19, indicating a dominance of unswitched states. The magnitude of the domain-wall length at the peak too varies depending on the bias voltage amplitude. For a specific $V_f$ = 4 V, the peak domain-wall length is higher than those observed under either lower $V_f$ = 3.8 V or higher $V_f$ = 15.0 V. The domain switching states under a lower bias voltage $V_f$ = 3.8 V are shown in the supplemental materials document. We point out here that the calculation of the domain wall length during switching of the domains was carried out by two formalisms - one by considering the nearest neighbor grids and another by considering next nearest neighbor grids. The results obtained from the second formalism are shown in the Fig. 13 while the supplemental materials document contains those obtained from the first formalism. The qualitative features of the results are similar in both the cases. The dependence of the domain wall length on the bias voltage amplitude and damping parameter $\rho$ could be clearly observed. Therefore, the phase-field simulation clearly shows that the domain-wall length maximizes under a specific bias voltage amplitude which triggers switching of domain over a specific timescale and allows completion of the nucleation and growth process. 

It is also possible to observe a correlation between the bias voltage pulse profile dependence of negative capacitance (and maximum negative capacitance at a specific voltage amplitude and timescale) and the domain switching mechanism. Switching of domains in a ferroelectric in presence of lattice defects has been modeled as a change in the shape and size of an elastic medium in presence of pinning. Using molecular dynamics simulation as well as experimental results, it has been shown \cite{Yang,Liu} that application of a bias across a wide amplitude and time scale range results in dynamic crossover in domain switching mechanisms - from relaxation to creep and depinning (when the bias voltage becomes comparable with the pinning potential) to finally, viscous flow as the voltage and frequency enhance. The domain wall velocity ($v_d$) versus bias electric field ($E$) follows $V_d$ $\propto$ $E^{\alpha}$ patterns with crossover in the exponent $\alpha$. We fitted the $V_C$ versus frequency ($f = 1/t$) data analogously (Figs. 3a and 3b) and observed crossover in domain switching mechanism from creep to creep + depinning. Interestingly, we observed maximum negative capacitance at a timescale which corresponds to this crossover point. The crossover point from creep to creep+depinning appears to have resulted in maximization of the domain-wall length during switching which, in turn, yields maximum transient negative capacitance.         

\section{Conclusion}
In conclusion, this work indicates a correlation among bias votage pulse profile (ampitude and timescale), domain-wall density during switching, domain switching mechanism, shape of the ferroelectric hysteresis loop, and transient negative capacitance. When the bias voltage pulse profile induces ferroelectric domain switching with maximization of domain-wall length during the ``switching" process, the specific negative capacitance maximizes. The corresponding ferroelectric hysteresis loop exhibits minimum radius of curvature at coercive voltage. The switching kinetics of the ferroelectric system is governed by both the free energy landscape associated with the structural transition as well as the extrinsic effects such as defect density driven nucleation and growth kinetics. This observation could be utilized to stabilize the ``optimum" domain-domain-wall configuration in a ferroelectric/dielectric heterostructure to maximize the static negative capacitance which would have important ramifications for negative capacitance based devices. In fact, it is possible to use the observed differential voltage amplification across the dielectric capacitor for specific applications. Although, the amplification too has been observed across a specific timescale, it can be utilized in different sensor circuits for improving the sensitivity across a timescale of importance. Moreover, since the timescale of observing maximum differential voltage amplification is related to the timescale of ferroelectric domain switching which, in turn, depends on the nucleation and growth kinetics, it is possible to tune the timescale by controlling the defect density, i.e., density of the nucleation centers. Therefore, tuning of timescale of observing maximum differential voltage amplification via defect density in the ferroelectric capacitor could open a new possibility of applications. Interestingly, it has been shown \cite{Graf} very recently that the voltage amplification could be associated with stronger electrostrictive effect in the dielectric capacitor. This, in turn, too could have application potential. It is also important to mention that since the epitaxial thin films of the most widely studied and used ferroelectric compound with two different crystallographic orientations containing different extent of epitaxial strain exhibit identical features repeatedly, these key features – outlined above – appear to be robust and applicable for all the proper ferroelectric systems. Verification in improper ferroelectric systems, of course, is necessary which will be taken up in near future. \\ 

\noindent $\textbf{SUPPLEMENTARY MATERIALS}$\\
The X-ray photoelectron spectroscopy (XPS) data, additional $P-V$ hysteresis loops recorded at different timescale, results of differential voltage amplification, domain switching kinetics measured under different bias voltage, polarization and leakage current versus bias voltage data, and the results of simulation of the domain switching kinetics under a lower bias voltage are available in the supplementary materials.\\

\noindent $\textbf{ACKNOWLEDGMENTS}$\\
This work is supported by Anusandhan National Research Foundation (ANRF), Government of India, project (SPR/2021/000131). One of the authors (GSK) acknowledges helpful discussion with M. Hoffmann, B.J. Rodriguez, P. Choudhury, and K. Chatterjee and also the financial support from DST-INSPIRE Fellowship of Government of India. Another author (DM) acknowledges the financial support received from the Technical Research Centre (TRC), Department of Science and Technology, Government of India (Grant No. AI/1/62/IACS/2015) and also from the India Russia Joint Research project of Department of Science and Technology, Government of India (Grant No. DST/IC/RSF/2024/542).\\

\noindent $\textbf{AUTHOR DECLARATION}$\\

\noindent $\textbf{Conflict of Interest}$\\

\noindent The authors have no interest to disclose.\\

\noindent $\textbf{Author contributions}$\\

\noindent $\textbf{Ganga S. Kumar}$: conceptualization (lead), data curation (lead), formal analysis (lead), investigation (lead), writing original draft (lead). $\textbf{Sudipta Goswami}$: data curation (supporting), formal analysis (supporting). $\textbf{Shubhasree Chatterjee}$: data curation (supporting), formal analysis (supporting). $\textbf{Dilruba Hasina}$: data curation (supporting), formal analysis (supporting). $\textbf{Miral Verma}$: software (lead), validation (lead). $\textbf{Devajyoti Mukherjee}$: funding acquisition (lead), data curation (supporting), formal analysis (supporting), supervision (lead), validation (lead). $\textbf{Chandan Kumar Ghosh}$: funding acquisition (lead), formal analysis (supporting), supervision (supporting), validation (supporting). $\textbf{Dipten Bhattacharya}$: conceptualization (lead), funding acquision (lead), project administration (lead), supervision (lead), validation (lead), writing - review and editing (lead).\\

\noindent $\textbf{Data Availability Statement}$\\
The data that support the findings of this study are available in the main paper, online supplementary materials document, and also upon reasonable request from the author.\\

\noindent $\textbf{REFERENCES}$

\end{document}